\documentclass[aps]{revtex4}
\def\BibTeX{{\rm B\kern-.05em{\sc i\kern-.025em b}\kern-.T
    08em\kern-.1667em\lower.7ex\hbox{E}\kern-.125emX}}
\usepackage{graphicx}

\usepackage[a4paper]{geometry}
\begin{document}

\title{Experimental realization of magnetic energy concentration and transmission at a distance by metamaterials}

\author{Jordi Prat-Camps, Carles Navau, and Alvaro Sanchez}
\affiliation{Departament de F\'{\i}sica, Universitat Aut\`onoma de Barcelona, 08193 Bellaterra,
Barcelona, Catalonia, Spain}

\begin{abstract}
Controlling electromagnetic energy is essential for an efficient and sustainable society. A key requirement is concentrating magnetic energy in a desired volume of space in order to either extract the energy to produce work or store it.  Metamaterials have opened new possibilities for controlling electromagnetic energy \cite{tjc_book,zheludev}. Recently, a superconductor-ferromagnetic metamaterial that allows unprecedented concentration and amplification of magnetic energy, and also its transmission at distance through free space, has been devised theoretically \cite{concentrator}. Here we design and build an actual version of the superconductor-ferromagnetic metamaterial and experimentally confirm these properties. We show that also a ferromagnetic metamaterial, without superconducting parts, can achieve concentration and transmission of energy with only a slight decrease in the performance. Transmission of magnetic energy at a distance by magnetic metamaterials may provide new ways of enhancing wireless power transmission, where efficiency depends critically on the magnetic coupling strength between source and receiver.
\end{abstract}

\maketitle

Electromagnetic fields power our society. Achievements such as the ubiquitous availability of electric power and the globally connected world through internet have required an exquisite control of electromagnetic fields - how to create, transport and use them- and the development of continuously improving materials. An essential part of this effort is in the control and manipulation of static magnetic fields. They play a fundamental role, for instance, in the generators that provide the energy for our appliances and devices, or in the writing heads of computer memories for storing information. 

Metamaterials recently joined in to open new possibilities in the control of electromagnetic fields \cite{pendry,controlling,review_TO,zheludev}. Novel ways of concentrating energy at different scales and wavelengths have been achieved by metamaterials. Examples range from electromagnetic energy concentration by plasmonics at the nanoscale 
\cite{schuller,aubry} to thermal energy concentration at larger scale \cite{NarayanaT,Han12}. Metamaterials have also created new possibilities of controlling static magnetic fields. For example, magnetic cloaks have been proposed and experimentally realized using superconductors combined with magnetic materials \cite{wood,magnus,ourAPL,antimagnet,narayana,gomory,carpet_magnetic}.

In this work we focus on concentration of magnetic energy. A particularly needed feature is to achieve large magnetic field values not only by close contact with the sources but also at a distant point separated by an air gap, as in wireless power transmission. Also, it is important that magnetic energy can be concentrated in an empty region, where an antenna o sensor can be placed.
With these goals in mind, we theoretically designed in \cite{concentrator} an infinite cylindrical shell with extreme radial and angular permeability components $\mu_{\rho} \rightarrow \infty$ and $\mu_{\theta} \rightarrow 0$, respectively. We showed that such a shell would concentrate an external applied field in its interior. It increases the field magnitude by a factor that is simply the ratio of outer to inner shell radii, which can be tuned to achieve large values.  The shell would also expel the field of an interior source to the exterior, increasing its magnitude. These two properties, fully analytically demonstrated, allowed to propose different combinations of shells to obtain unprecedented control of magnetic fields. For example, fields could be concentrated at points distant from the source through free space, which could be relevant to enhance wireless transmission of power.

The ideal concentrating shell required extreme magnetic properties, not directly found in natural materials. To solve this, we proposed magnetic metamaterials made of alternated superconductor (SC) and ferromagnetic (FM) pieces to approximate the required anisotropic permeabilities. These pieces could be wedges or rectangular prisms 
\cite{concentrator}. Theoretical calculations indicated that a good performance should be obtained by metamaterial shells made of alternated wedge-shaped FM and SC pieces and behavior would improve with the number of elements \cite{sust}. However, ideal materials with linear behaviour and with very large permeability for the FMs and zero permeability for the SCs were assumed. Also, only infinitely long geometries were considered in the calculations. Thus, an experimental validation with actual materials and realistic geometries is needed to confirm these ideas.

In this work we design and construct actual metamaterial shells to experimentally confirm the properties of magnetic concentration, expulsion, and concentration at a distance. We will find that all these properties are achieved when using existing (not mathematically ideal) SC and FM materials, when having shells with short lengths, and when discretizing the continuous ideal material into a set of rectangular prisms. As an important step towards the feasibility of implementing our ideas in the technology, we will also experimentally demonstrate that metamaterial shells composed of only FMs separated with air gaps (without the use of SCs and their associated cryogenics) show good properties as well. Finite-element numerical calculations will simulate and help to interpret the experimental results.

We build a metamaterial cylindrical shell made of 36 rectangular prisms of alternated SC and FM materials (see Fig. 1a and Appendix for details).
The FM pieces are made of a commercial high-permeability metallic alloy (mu-metal) whilst the SC ones of a commercial coated conductor. The pieces are radially distributed and fixed in a non-magnetic plastic support specially designed and made by a 3-D printer (Fig. 1b,c). 


We first experimentally study the magnetic concentration properties of our metamaterial shell. Two Helmholtz coils created a uniform external field ($B_a$) perpendicular to the shell axis. A Hall probe was used to measure the concentrated field ($B^{\rm IN}_z$) in the central point inside the shell (see the sketch of the experimental setup in Fig. 2a and Appendix). Measurements were first performed with both the metamaterial and the probe submerged in liquid nitrogen, below the critical temperature of the SC, $T_c$. The magnetic-field enhancement factor at the center for the ideal infinitely long metamaterial should be 4, corresponding the ratio between outer to inner shell radii \cite{concentrator}.
Measurements  below $T_c$ showed that field was increased by a factor $2.70$ with respect to applied field (green symbols and dashed fitting line in Fig. 2b). This number is in excellent agreement with the 3D numerical simulation of this system (green solid line and Fig. 2d), which considered the actual finite dimensions and assumed ideal magnetic properties of the materials ($\mu_{FM}\to \infty$ and $\mu_{SC}\to 0$) . Simulation also shows that, despite the finite shell length, magnetic field achieves a very homogenous concentration \cite{concentrator}, which we experimentally confirmed.

We next study if similar concentration results can be produced with simplified versions of the metamaterial. Because using SCs requires cryogenics and this may limit the applicability of the device, we measured the concentration of magnetic field at temperatures above $T_c$, at which the SCs are deactivated. A concentrating factor of $2.23$ (red symbols and dashed fitting line in Fig. 2b) was still achieved under these conditions. The corresponding simulation with only 18 FM pieces (red solid line and Fig. 2e) perfectly matched these measurements. 
Simulations also illustrate the role played by the two different materials. The SC layers in the SC-FM metamaterial (Fig. 2d) prevent angular components of magnetic field in the shell volume, so that magnetic field is basically guided radially to the central hole by the FM pieces. In the only-FM metamaterial shell (Fig. 2e), this role played by the SC parts is now taken up by the air gaps. This works only approximately, as can be seen from the presence of some field in the inner parts of the FM-metamaterial shell volume (Fig. 2e).

We next experimentally study the expelling properties of our metamaterial shell by measuring the field at its exterior when a dipolar-like magnetic field source is present in the shell interior. A small coil was placed at the center of the shell with its axis on the $z$ direction (see Fig. 3a) and was fed with a constant current. The field $B_z$ was measured outside the shell along the centered line as a function of the distance $z$. Measurements below $T_c$ (green symbols in Fig. 3b) show the expelled field was increased by a factor of about 2.4 in all the positions with respect to the measurements of the bare coil (blue symbols). Above $T_c$, the shell also increased the field in all exterior points by a factor of about 2, confirming its good behavior even without SCs. 
These measurements agree well with the corresponding 3D simulations performed assuming ideal materials (solid lines in Fig. 3b and Fig. 3c-e). Plots of the field at median planes show how the metamaterial shell expels most of the field from its interior to the exterior (Fig. 3d) as already discussed for the concentration case. When using solely FM pieces (Fig. 3e) the field expulsion is reduced.

The combination of the expulsion and concentration properties can be used to transfer magnetic energy at a distance, as theoretically anticipated in 
\cite{concentrator}. To experimentally demonstrate this property we used the studied shell with a coil inside as a source and construct a second shell as a receiver (see Figs. 4a and b). The second shell is an only-FM metamaterial shell. It consists of 18 FM pieces, without superconducting parts. We separate both shells at a distance of 70mm from center to center, so there is an air gap of 10mm between them. Measurements are shown in Fig. 4c, for two cases. When $T<T_c$ (upper plot), the shell with the source behaves like a full SC-FM metamaterial and, even though the receiver shell is a only-FM metamaterial shell, the measured magnetic field inside the receiver is increased by an average factor of around 7.5 with respect  to the field of the naked coil.
When $T>T_c$ (lower plot), the average factor is still around 6. In both cases it is seen that not only the magnetic field is magnified in the inner region of the receiver shell but also the field gradient is enlarged, as predicted by the theory \cite{concentrator}.
The main result here is that these experiments are a proof-of-principle of the theory for magnetic energy transmission with metamaterials. The particular obtained numbers can be further optimized. The field-increase factor of 7.5 or 6 and also the air gap distance, can be both easily turned into larger values by changing the geometry, like reducing the inner radii of both shells \cite{concentrator}. Moreover, modifications of the studied magnetic metamaterials can be explored for changing the magnetic field shape in the space around the shell as well.

When a second coil is placed in the receiver shell, the enhanced field transmission is equivalent to increasing the magnetic coupling between both coils, which may have applications in wireless transmission of power \cite{persp,Kim13}.  In wireless systems, including  the recent non-radiative power transfer proposals \cite{kurs}, the mutual inductive coupling between the source and the receive
resonators is a key parameter to increase the efficiency. As discussed in \cite{DaHuang}, because the distance between the source and receiver
is so much smaller than the wavelength, the relevant
field distribution is quasistatic and the inductive coupling
relates predominantly to magnetic flux emanating
from one coil captured by the second coil. This is precisely what we achieve with our metamaterial shells. Actually, magnetic metamaterials have been already proposed for increasing the magnetic coupling in wireless systems in the form of magnetic lenses. This strategy requires a negative permeability material \cite{choi,urzhumov,DaHuang,mitsu}, whereas our metamaterial is made of non-resonant materials with positive permeability. 

Even though our ideas have been confirmed strictly in the static limit, there are good indications that for low frequencies (up to tenths of kilohertz at least) the materials will behave as required. Recently, a magnetic cloak made of similar FM and SC materials designed to operate at the dc regime was shown to maintain cloaking properties for low-frequency applied magnetic fields \cite{accloak}. For wireless power applications, further work should be done to study the performance of these magnetic metamaterials at higher frequencies, including eventual losses in the components.

 To sum up, we have experimentally demonstrated the unique properties that superconductor-ferromagnetic metamaterials can offer in terms of magnetic energy concentration, expulsion, and transmission through free space. Although the original theory was derived for infinitely samples and assumed ideal materials, our results confirm that these properties can be achieved in finite geometries and using commercially available materials. Finite-element calculations precisely reproduce the measured data and thus can be used to obtain optimized designs. Even when the superconducting parts are removed, very good behaviour is obtained with only ferromagnetic metamaterials. This hints at possible application of the results to technology, in particular in increasing the magnetic coupling between distant circuits, an essential factor for enhancing wireless transmission of power.   

\section*{Acknowledgements}

We thank Elena Bartolome and EUSS for support, Fedor Gomory and Jan Souc for helpful discussions, SuperPower for providing the coated conductors, and Spanish Consolider Project NANOSELECT (CSD2007-00041) and MAT2012-
35370 for financial support. JPC acknowledges a FPU grant form Spanish Government
(AP2010-2556).





\section*{Corresponding author}

Alvaro Sanchez (alvar.sanchez@uab.cat)

\newpage

\begin{figure}[T]
	\centering
		\includegraphics[width=1.0\textwidth]{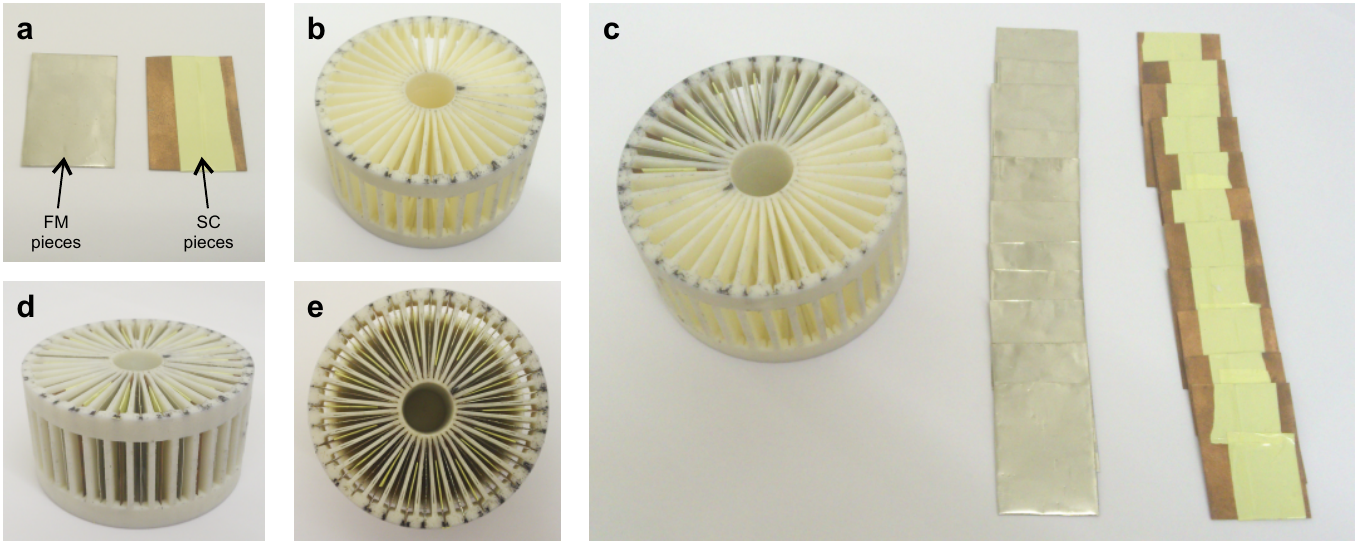}
	\caption{
The metamaterial shell was made of 18 pieces of FM material and 18 of SC material, {\bf a}, which were radially and alternately displaced. To correctly position these pieces a non-magnetic support structure, {\bf b}, was fabricated. This plastic structure was 3D-printed having 36 lodgings homogeneously separated in which the SC and FM pieces fit, {\bf c}. The resulting metamaterial shell, {\bf d-e}, had an approximate interior and exterior radii of $R_1=7.5$mm and $R_2=30$mm respectively, and a height of 30mm. \vspace{10mm}}
	\label{fig1}
\end{figure}

\begin{figure}[T]
	\centering
		\includegraphics[width=1.0\textwidth]{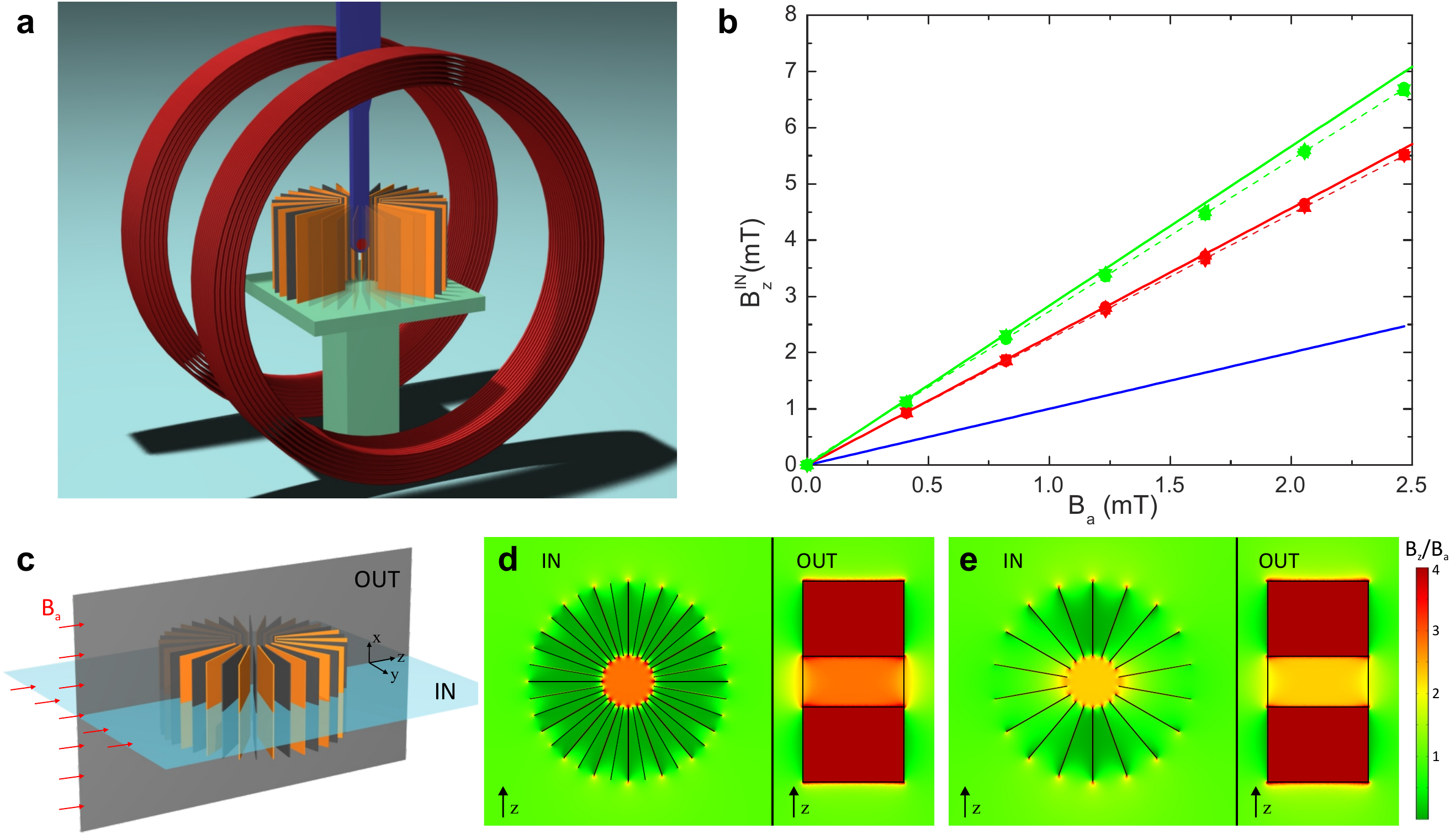}
	\caption{
{\bf a} Sketch of the experimental setup used to measure the magnetic field concentration. Two Helmhotlz coils (red) with a radius of 70mm were used to create a uniform field. The metamaterial shell was placed in the central region (SC pieces represented in orange and FM ones in dark gray) and the Hall probe (blue) measured the field in the central interior point (some pieces have been sketched  translucent to show the precise measuring point). {\bf b} The measured field inside the metamaterial shell is plotted as a function of the field applied by the coils $B_a$ (in symbols) and for two different working conditions; measurements for $T<T_c$ in green and for $T>T_c$ in red (horizontal and vertical error bars $\Delta B_a=0.04$mT and $\Delta B^{\rm IN}=0.002$mT respectively, have been omitted for clarity). The corresponding linear fits are presented in dashed lines (whose slopes are 2.70 and 2.23 for $T<T_c$ and  $T>T_c$, respectively), together with the simulations results in solid lines (whose slopes are 2.83 and 2.28 for $T<T_c$ and  $T>T_c$, respectively), showing a good agreement. The blue line shows the field created by the bare coils in the central point, and is relevant to show the improvement achieved by the metamaterial shell in both working conditions. 
The calculated $B_z$ field component is plotted in the planes IN and OUT, {\bf c}, for the case of 18 SC alternated with 18 FM (i.e. $T<T_c$), {\bf d}, and for the case of only 18 FM ($T>T_c$), {\bf e}.}
	\label{fig2}
\end{figure}

\begin{figure*}[t]
	\centering
		\includegraphics[width=1.0\textwidth]{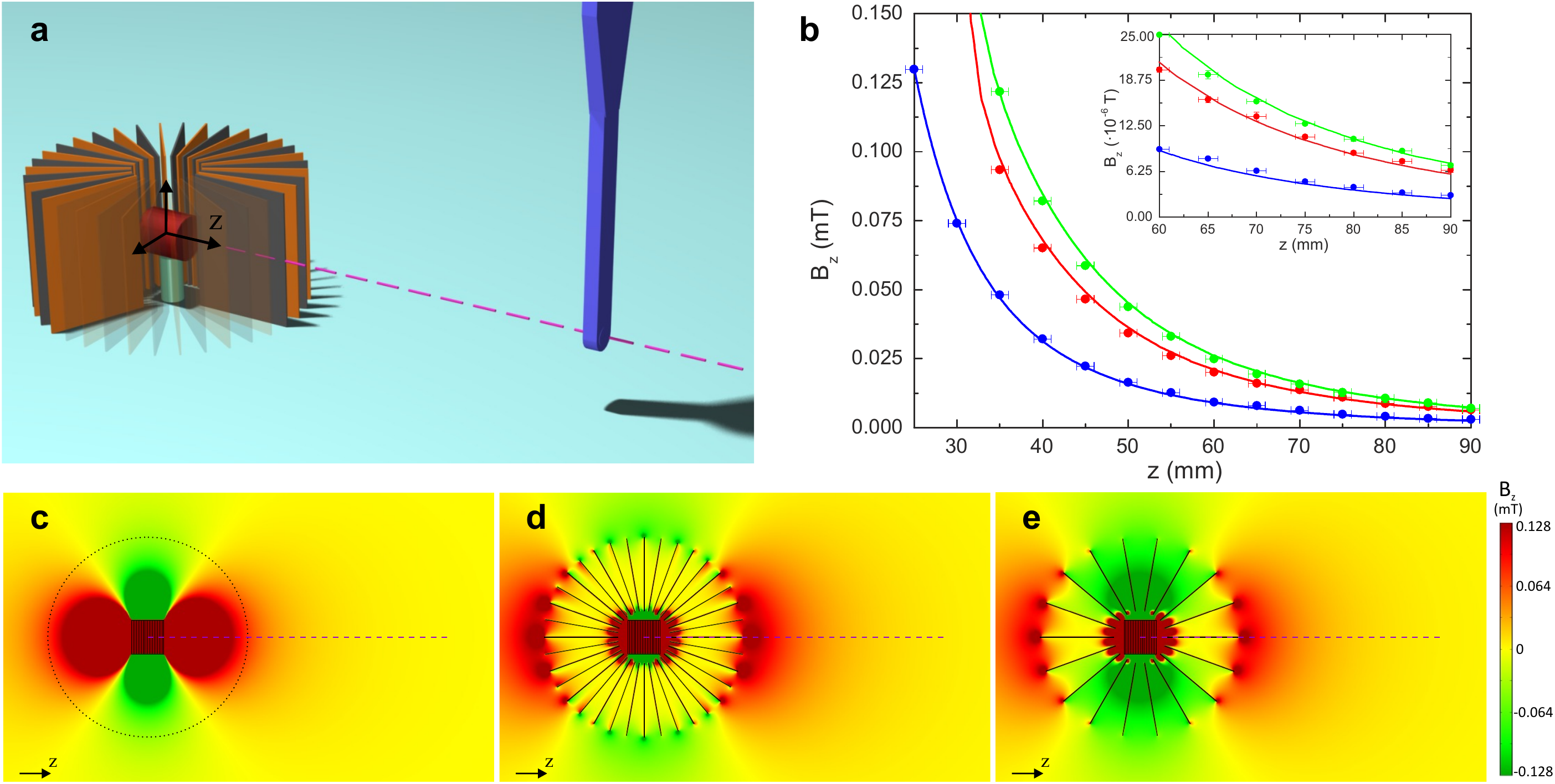}
	\caption{
{\bf a} Sketch of the experimental setup used to measure the expulsion capabilities of the shell. A small coil (red) was placed inside the metamaterial shell and was fed with a DC current. The $B_z$ field component was measured outside the shell by the Hall probe along the centered line (pink) as a function of the distance to the center $z$. {\bf b} Measurement results are plotted in symbols; blue symbols correspond to the field of the isolated coil, red symbols are measurements outside the shell at $T>T_c$ and green ones when the measurements were performed at $T<T_c$. Solid lines show the results of the corresponding 3D simulations, in full agreement with the measurements. {\bf c-e} Plots of the $B_z$ field component in the median plane of these simulations for the case of the coil alone,{\bf c}, 18 FM and 18 SC ($T<T_c$), {\bf d}, and 18 FM ($T>T_c$), {\bf e}.}
	\label{fig3}
\end{figure*}

\begin{figure*}[t]
	\centering
		\includegraphics[width=1.0\textwidth]{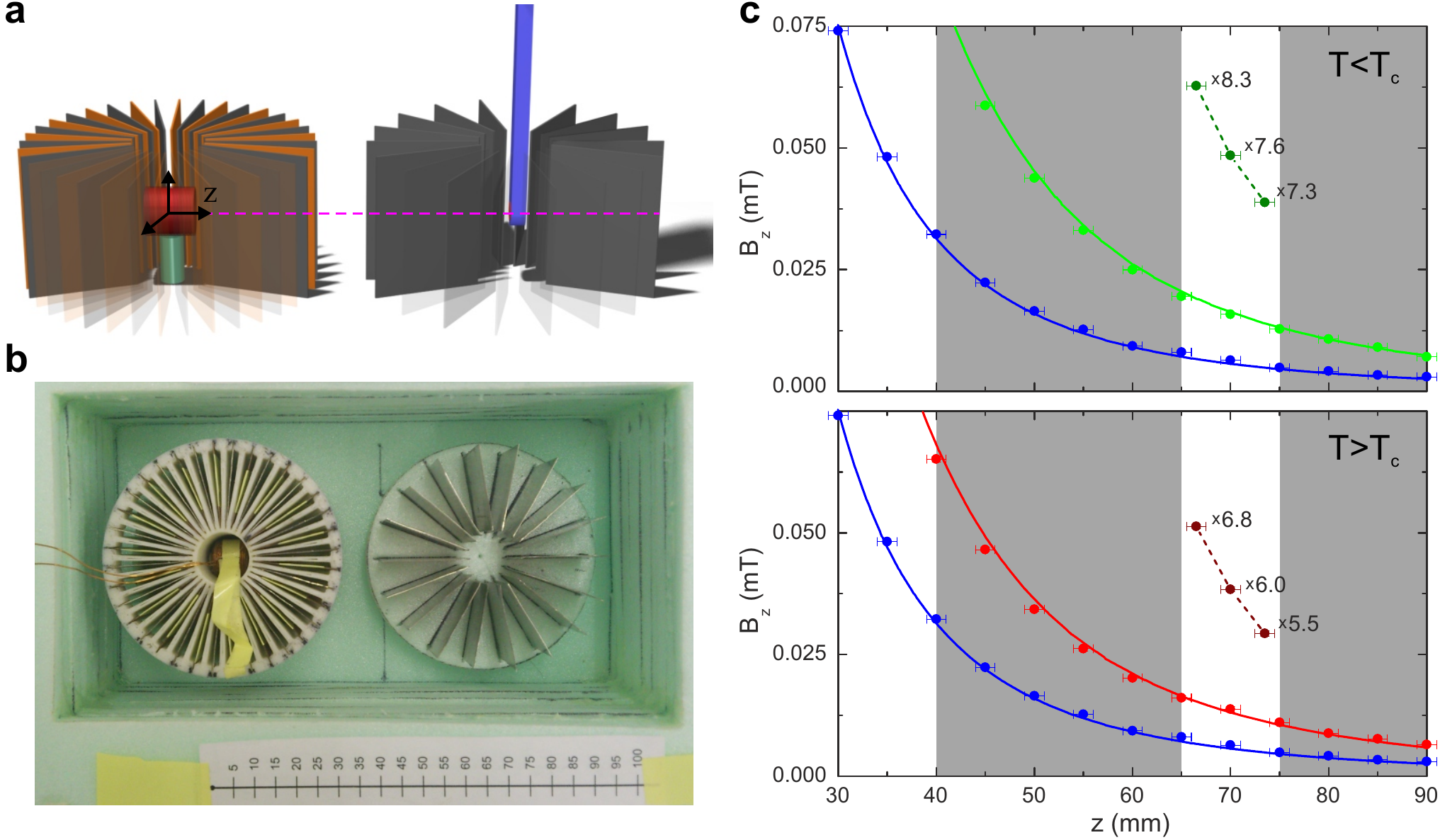}
	\caption{
{\bf a} Sketch of the experimental setup for magnetic field concentration at a distance. The coil (red) was placed inside the metamaterial shell. A second metamaterial shell with a similar size made of only 18 FM pieces was placed at a distance of 70mm from the center of the first shell. The field was measured by the Hall probe inside the second shell aligned with the coil. {\bf b} Picture of the real experimental setup, with the coil placed inside the original metamaterial shell (left), and the only-FM metamaterial shell (right). 
{\bf c} Measurement results (symbols) for the two working conditions; $T<T_c$ (upper plot) and $T>T_c$ (lower plot). In the upper plot, blue symbols and line correspond to measurements and simulation of the coil alone, and light green symbols and line correspond to measurements and simulation of the field expelled by the metamaterial shell at $T<T_c$ (these results were already presented in the previous figure and are only included for comparison). The dark green symbols correspond to the field measured inside the second shell when both shells are present (dashed line is a guide for the eye). The numbers next to the symbols indicate the enhancement factor of the measured field with respect to the field created by the bare coil at the same positions. In the lower plot, blue symbols and line correspond to the case of the bare coil, red symbols and line represent the expelling measurements and simulation at $T>T_c$, and dark red symbols correspond to the field measured inside the second shell when the two shells are present (dashed line is a guide for the eye). The enhancement factors are also shown.}
	\label{fig4}
\end{figure*}

\section*{APPENDIX}

\subsection*{Metamaterial shell construction}
The concentrating metamaterial shell consisted of 18 FM pieces and 18 SC pieces with a rectangular shape of 22x30mm (See Fig. 1). The former were made of a commercial mu-metal foil with a thickness of $0.3$mm from Vacuumschmelze, which was cut to the appropriate size. The SC parts were made of type-II SC strip 12mm width (SuperPower SCS12050). Each part was made of two pieces of strip, cut with a length of 30mm and parallel fixed with adhesive tape with an approximate overlap of 2mm.

To ensure the appropriate placement and alignment of the SC and FM pieces, a non-magnetic plastic support was designed and built. The support had a hollowed cylindrical shape, with an external radius of 31mm, an interior one of 6.5mm and a height of 33mm and had 36 lodgings homogeneously distributed with a consecutive angular spacing of $10^{\circ}$. It was provided with openings in the lateral face and also in the interior walls to ensure the free circulation of the liquid nitrogen during the measurements in which it was required. The support was built with a 3D printer and was made of ABS thermoplastic polymer. 

\subsection*{Concentration measurements and simulations}
The Helmholtz coils used to generate the uniform applied field had a radius of 70mm and were aligned and separated a distance of 70mm. We used a Hall probe model HHP-NP from Arepoc to measure the magnetic fields. Measurements below $T_c$ were done by submerging the metamaterial shell and the Hall probe into liquid nitrogen. The measurements where done increasing the intensity in the coils from 0 to 0.6A and measuring the field in the shell hole. Measurements were repeated several times for each configuration. 


\subsection*{Expulsion measurements and simulations}
The small coil used to generate the field had an approximate diameter of 10mm and a height of 10mm and was vertically centered in the interior hole of the metamaterial shell. The field was measured by placing the Hall probe at the same height outside the shell. Measurements at $T>T_c$ were performed by submerging the metamaterial shell, the coil and the Hall probe into liquid nitrogen. 
Measurements were done increasing the field in the coil from 0A to 1A, and measuring the corresponding field. The transmitted field was calculated as the slope of the resulting field-intensity plot, which allowed to separate it from the ambient fields and the field created by the remanent magnetization of the ferromagnets. This process was repeated mutiple times in each measuring position, and in all the cases the plots field-intensity showed a very linear behavior, with no sign of saturation. 
In the simulations, the coil was modeled as a uniformly magnetized cylinder with the same size of the real coil. The magnetization was tunned to match the calculated plot of the coil alone (blue solid line) with the corresponding experimental measurements (blue symbols). SC and FM pieces were assumed ideal.

\subsection*{Concentration at distance measurements and construction of a only-FM shell}
For these measurements we built a second metamaterial shell consisting of 18 FM pieces (made of the same mu-metal foil) with a rectangular shape similar to that of the first shell. The pieces where homogeneously distributed with an angular spacing of $20^{\circ}$ and where fixed in a non-magnetic extruded polystyrene foam layer in which 18 lodgings where made. The resulting shell had an approximate interior and exterior radii of $R_1=7.5$mm and $R_2=30$ respectively, and a height of 30mm.
The field was measured inside the second shell for different positions $z$ using the procedure previously described, and for temperatures above and below $T_c$.

\subsection*{Numerical simulations}
3D simulations where obtained by the AC/DC module of Comsol Multiphysics software assuming ideal materials ($\mu^{SC}=10^{-5}$ and $\mu^{FM}=10^{5}$). No free parameters were used for the calculations.


\begin{thebibliography}{99}


\bibitem{tjc_book} Cui T. J.,  Smith, D. R., and Liu, R. 
Metamaterials: Theory, Design and Applications (New York: Springer) (2010).
		
\bibitem{zheludev}
Zheludev, N. I.	and Kivshar, Y. S. From metamaterials to metadevices.
\textit{Nature Materials} {\bf 11}, 917 (2012).
    
		\bibitem{concentrator} 
Navau, C., Prat-Camps, J. \& Sanchez, A.
Magnetic energy harvesting and concentration at a distance by transformation optics.
\textit{Phys. Rev. Lett.} {\bf 109}, 263903 (2012).

\bibitem{pendry} Pendry, J. B., Holden, A. J., Robbins, D. J., and Stewart, W. J.
Magnetism from conductors and enhanced nonlinear phenomena.
IEEE Trans. on Microwave Theory and Techniques {\bf 47}, 2075 (1999).  



\bibitem{controlling} Pendry,J. B. , Schurig, D., and Smith, D. R. 
Controlling electromagnetic fields. 
\textit{Science} 312, 1780 (2006).

\bibitem{review_TO} Chen H., Chan C. T., and Sheng, P. 
Transformation optics and metamaterials. 
\textit{Nature Materials} {\bf 9}, 387 (2010).









\bibitem{schuller} Schuller J. A., Barnard E. S., Cai W., Jun Y. C., White, J. S., and Brongersma, M. L. 
Plasmonics for extreme light concentration and manipulation.
\textit{Nature Materials} {\bf 9}, 193 (2010).


\bibitem{aubry} Aubry, A. et al., 
Plasmonic Light-Harvesting Devices over the Whole Visible Spectrum. 
\textit{Nano Letters} {\bf 10}, 2574 (2010).




\bibitem{NarayanaT} Narayana, S. and Sato, Y.
Heat Flux Manipulation with Engineered Thermal Materials. 
\textit{Phys. Rev. Lett.} {\bf108}, 214303, (2012).




\bibitem{Han12} Han, T., Zhao, J., Yuan, T. Lei, D., Li, B., and Qiu, C.-W.
Theoretical realization of an ultra-efficient thermal- energy harvesting cell made of natural materials. 
\textit{Energy Environ. Sci.} DOI: 10.1039/c3ee41512k (2013).



\bibitem {wood} Wood, B. \& Pendry, J. B. 
Metamaterials at zero frequency. 
\textit{J. Phys. Condens. Matter} 19, 076208 (2007).


\bibitem{magnus} Magnus, F. et al. 
A d.c. magnetic metamaterial. 
\textit{Nature Materials} {\bf 7}, 295  (2008).


\bibitem{ourAPL} Navau, C., Chen, D.-X., Sanchez, A. and Del-Valle, N. 
Magnetic properties of a dc metamaterial consisting of parallel square superconducting thin plates. 
\textit{Appl. Phys. Lett.} {\bf 94}, 242501 (2009).


\bibitem{antimagnet} Sanchez, A., Navau, C., Prat-Camps, J., and Chen, D.-X. 
Antimagnets: controlling magnetic fields with superconductor–metamaterial hybrids. 
\textit{New J. Phys.} {\bf 13}, 093034 (2011).


\bibitem{narayana} Narayana, S. and Sato, Y. 
DC Magnetic Cloak. 
\textit{Advanced Materials} {\bf 24}, 71 (2012).

\bibitem{gomory}
Gomory, F., Solovyov, M., Souc, J., Navau, C., Prat-Camps, J. and Sanchez, A.
Experimental realization of a magnetic cloak.
\textit{Science} {\bf 335}, 1466 (2012). 

\bibitem{carpet_magnetic} Wang R., Mei, Z. L., and Cui, T. J.
A carpet cloak for static magnetic field.
\textit{Appl. Phys. Lett.} {\bf 102}, 213501 (2013).












\bibitem{sust} 
Prat-Camps, J., Sanchez, A., and Navau, C.
Superconductor-ferromagnetic metamaterials for magnetic cloaking and concentration.
\textit{Supercond. Sci. Technol.} {\bf 26}, 074001 (2013).


\bibitem{persp} Stewart, S.
The Power to Set You Free.
\textit{Science} {\bf 317}, 55 (2007).


\bibitem{Kim13} Kim, S., Ho, J. S., and Poon, A. S. Y.
Midfield Wireless Powering of Subwavelength Autonomous Devices. 
\textit{Phys. Rev. Lett.} {\bf110}, 203905, (2013). 




\bibitem{kurs} Kurs, A., Karalis, A., Moffatt, R., Joannopoulos, J. D., Fisher, P., and Soljacic, M.
Wireless Power Transfer via Strongly Coupled Magnetic Resonances
\textit{Science} {\bf 317}, 83 (2007).

\bibitem{DaHuang} 
Huang, D., Urzhumov, Y., Smith, D. R., Teo, K. H., and Zhang, J.
Magnetic superlens-enhanced inductive coupling for wireless power transfer.
\textit{J. Appl. Phys.} {\bf 111}, 064902 (2012).

\bibitem{choi} Choi, J. and Seo, C. 
High-efficiency wireless energy transmission using magnetic resonance based on negative refractive index material. 
\textit{Progr. Electromagn. Res.} {\bf 106}, 33 (2010).

\bibitem{urzhumov} Urzhumov, Y. and Smith, D. R.
Metamaterial-enhanced coupling between magnetic dipoles for efficient wireless power transfer
\textit{Phys. Rev. B} {\bf 83}, 31 (2011).

\bibitem{mitsu} Wang, B., Yerazunis, W., and Teo, K. H.
Wireless Power Transfer:
Metamaterials and Array
of Coupled Resonators
\textit{Proceedings of the IEEE} {\bf 101}, 1359 (2013).
| 

\bibitem{accloak} 
Souc, J., Solovyov, M., Gomory, F., Prat-Camps, J., Navau, C., and Sanchez, A.
A quasistatic magnetic cloak.
\textit{New Journal of Physics} {\bf 15}, 053019 (2013).

\end{thebibliography}
\end{document}